\documentclass[a4paper]{jpconf}
\usepackage{graphicx} 
\usepackage{subfigure}
\RequirePackage{lineno}
\begin{document}
\title{The physics mechanisms of light and heavy flavor $v_{2}$ and mass ordering in AMPT}

\author{Hanlin Li$^{1,2}$, Zi-Wei Lin$^{3}$, Fuqiang Wang$^{2}$}
\address{$^{1}$College of Science, Wuhan University of Science and Technology, Wuhan, Hubei 430065, China }
\address{$^{2}$Department of Physics and Astronomy, Purdue University, West Lafayette, IN 47907, USA}
\address{$^{3}$Department of Physics, East Carolina University, Greenville, NC 27858, USA}

\begin{abstract}
A Multi-Phase Transport (AMPT) model has been shown to describe experimental data
well, such as the bulk properties of particle spectra and elliptic anisotropy ($v_{2}$) in
heavy ion collisions. Recent studies have shown that AMPT describes the $v_{2}$ data in small
system collisions as well. In these proceedings, we first investigate the origin of the mass ordering of identified hadrons $v_{2}$  in heavy ion as well as small system collisions. We then study the production mechanism of the charm $v_{2}$ in light of the escape mechanism for the light quark $v_{2}$.
\end{abstract}
\section{Introduction\label{sec:intro}}
\quad 
Relativistic heavy ion collisions aim to create the quark-gluon plasma (QGP) and study its properties at the extreme conditions 
of high temperature and energy density~\cite{Arsene:2004fa,Back:2004je}. The collective flow as a soft probe
is often used to study the QGP properties in experimental and theoretical investigations. Both hydrodynamics~\cite{Heinz:2013th,SongMultistrange}and transport theory~\cite{Lin:2004en}can describe the bulk data in heavy ion collisions. For example, the string melting version of A Multi-Phase Transport (AMPT) model~\cite{Lin:2004en,Lin:2001zk} reasonably reproduces particle yields, $p_{\perp}$ spectra, and $v_2$ of low-$p_{\perp}$ pions and kaons in central and mid-central Au+Au collisions at 200A GeV and Pb+Pb collisions at 2760A GeV~\cite{Lin:2014tya}. The small system data can be also satisfactorily described by AMPT~\cite{Bzdak:2014dia}.
 
Recent studies have shown that light parton $v_{2}$ is mainly generated by the anisotropic parton escape from the collision zone and hydrodynamics may play only a minor role~\cite{He:2015hfa,Lin:2015ucn}. It suggests that the mass ordering of 
$v_{2}$, commonly believed as a signature of hydrodynamics, may arise from other mechanisms. In these proceedings, we first investigate the physic mechanisms of mass ordering
in AMPT~\cite{Li:2016flp,Li:2016ubw}. We then study the production mechanism of the charm $v_{2}$ in light of the escape mechanism for the light quark $v_{2}$.
\section{Model details and analysis method  \label{sec:setup1}}
\quad
We employ the string melting version of AMPT~\cite{Lin:2004en,Lin:2001zk} in our study. The model consists of four components:
fluctuating initial conditions, parton elastic scatterings, quark coalescence for hadronization, and hadronic interactions. The partons interaction are modeled by Zhang's parton cascade (ZPC)~\cite{Zhang:1997ej}. We use Debye screened differential cross-section $d\sigma/dt\propto\alpha_s^2/(t-\mu_D^2)^2$~\cite{Lin:2004en}, with strong coupling constant $\alpha_s=0.33$ and Debye screening mass $\mu_D=2.265$/fm (the total cross section is then $\sigma=3$~mb) for all AMPT simulation in our work. After partons stop interaction, a simple quark coalescence model is applied to combine two nearest partons into a meson and three nearest partons into a baryon(or antibaryon). The hadronic interactions are described by the ART model~\cite{B:art}. We terminate the hadronic interactions at a cutoff time. 

We simulate three collision systems: {Au+Au} collisions with {$b=6.6$-8.1~fm} 
at the nucleon-nucleon center-of-mass energy ${\sqrt{s_{_{\rm NN}}}}=200$~GeV, {$d$+Au} collisions with {$b=0$~fm} at ${\sqrt{s_{_{\rm NN}}}}=200$~GeV, and {$p$+Pb} collisions with {$b=0$~fm} at ${\sqrt{s_{_{\rm NN}}}}=5$~TeV by AMPT. We analyze the momentum-space azimuthal anisotropy of partons in the final state before hadronization, of hadrons right after hadronization before hadronic rescatterings take place, and of freeze-out hadrons in the final state. The momentum anisotropy is characterized by Fourier coefficients~\cite{Voloshin:1994mz}
\begin{equation}
v_2=\langle \cos 2(\phi -\psi_2^{(r)}) \rangle,
\end{equation}
where $\phi$ is the azimuthal angle of the particle (parton or hadron) momentum. The  $\psi_2^{(r)}$ is the harmonic plane of each event from its initial configuration of all partons.
All results shown are for particles within pseudo-rapidity window $|\eta|<1$.
\begin{figure}[h]
\begin{minipage}[t]{18pc}
\includegraphics[width=0.85\linewidth,height=7cm,clip=]{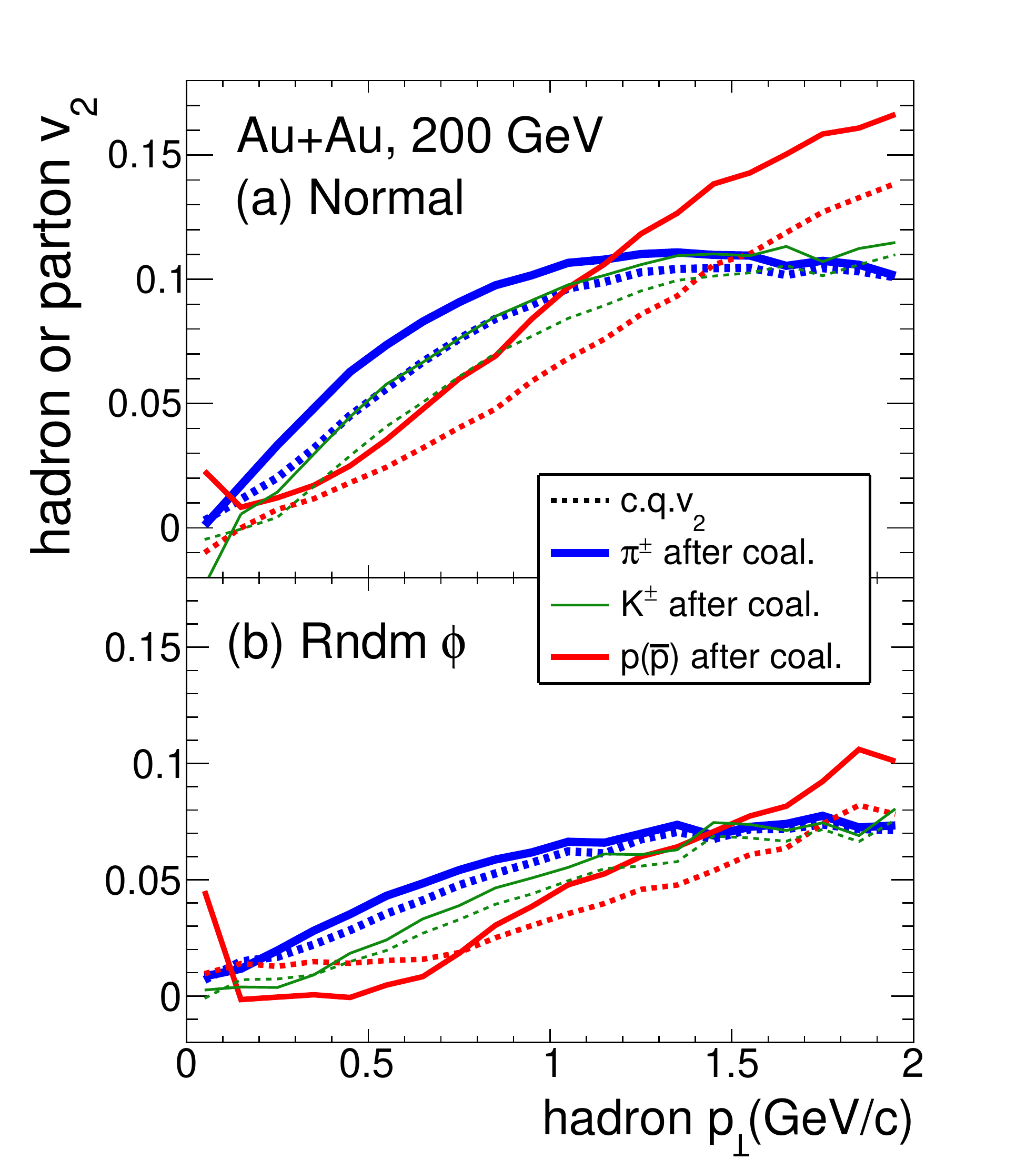}
\caption{\label{fig:fig1}Mass splitting from coalescence. Constitute quark (c.q., dashed curves) and primordial hadron (solid curves) $v_2$ both as a function of hadron $p_{\perp}$ in {Au+Au} collisions by normal (a) and $\phi$-randomized AMPT (b).}
\end{minipage}\hspace{2pc}%
\begin{minipage}[t]{18pc}
\includegraphics[width=0.85\linewidth,height=7cm,clip=]{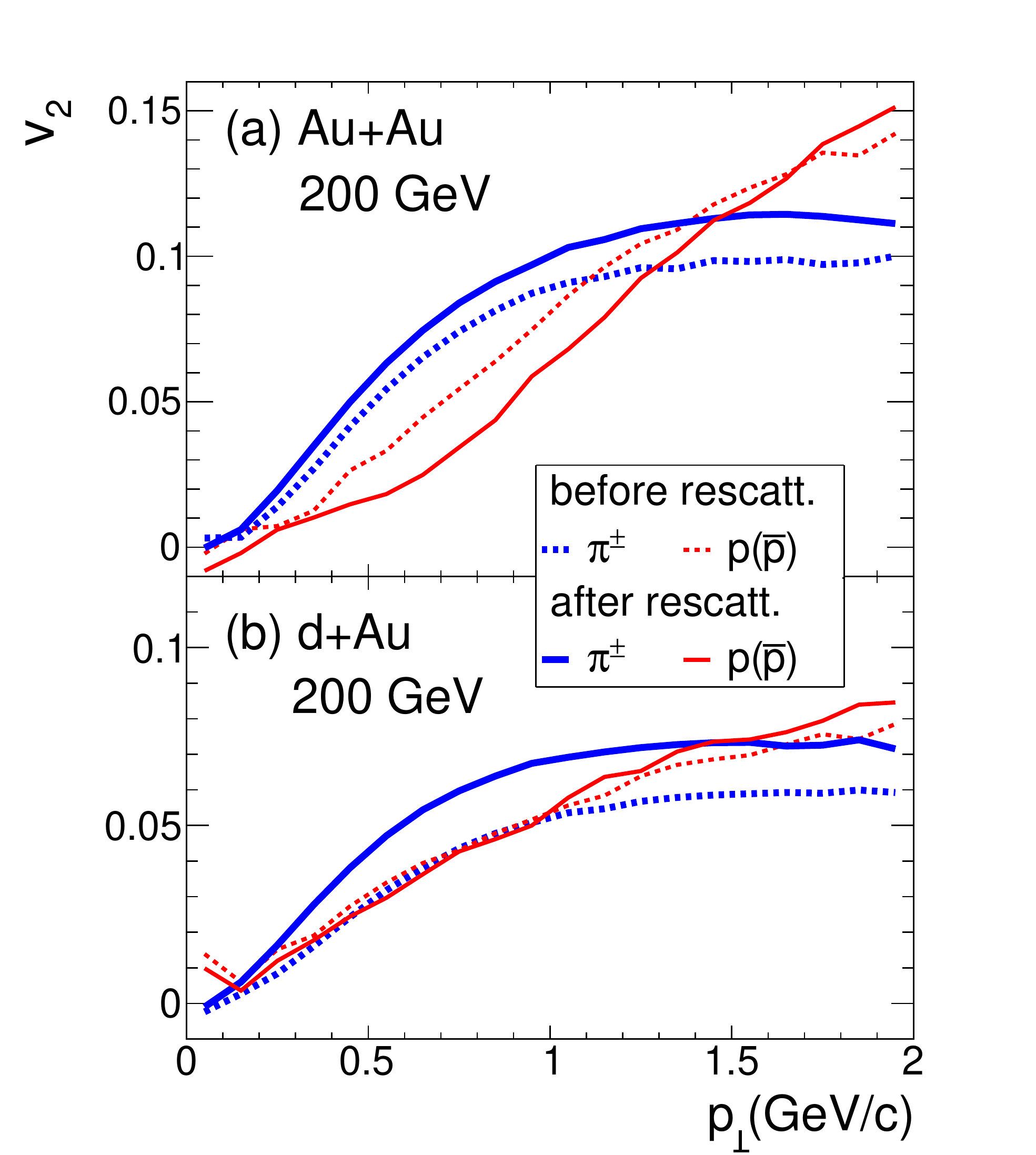}
\caption{\label{fig:fig2} Effects of hadronic rescattering. Pion and (anti) proton $v_2$ before (dashed) and after (solid) hadronic rescatterings in {Au+Au} (a) and
{$d$+Au} (b) collisions by AMPT.}
\end{minipage} 
\end{figure}
\section{Mass ordering of $v_2$ \label{sec:setup}}
\quad 
\begin{figure}[tbph]
\centering
 \includegraphics[width=30pc]{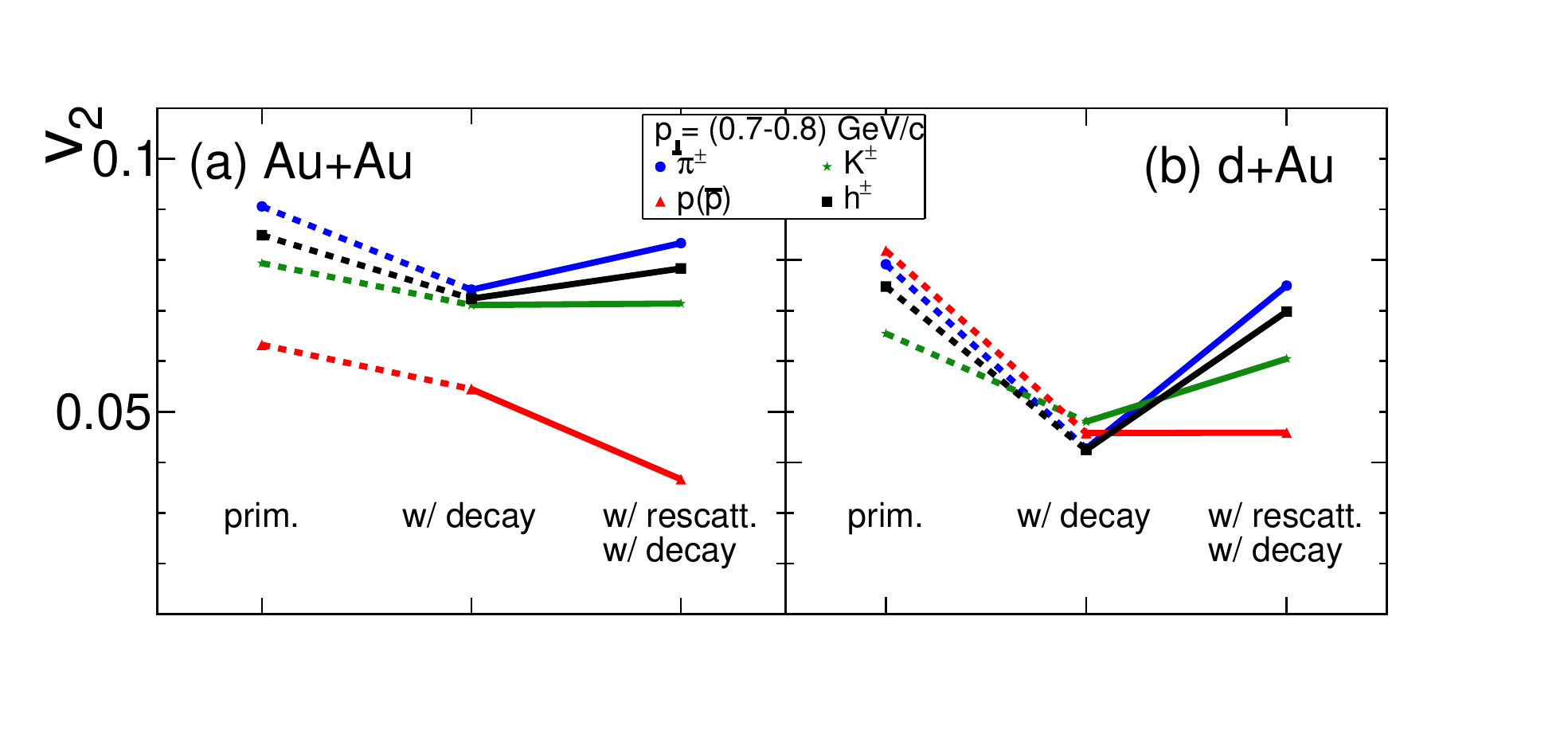}
 \caption{(Color online) Mid-rapidity ($|\eta|<1$) $v_2$ of $\pi$, K, p($\bar{p}$) and charged hadrons ($h^\pm$) at $0.7<p_{\perp}<0.8$~ GeV/$c$ at different stages of system evolution in $b=6.6$-8.1~fm Au+Au (left) and $b=0$~fm {$d$+Au} (right) collisions at ${\sqrt{s_{_{\rm NN}}}}$=200~GeV by AMPT: right after coalescence hadronization and before hadronic rescattering (initial $v_2$ of primordial hadrons), hadron initial $v_2$ including decays, and after hadronic rescattering at freezeout (hadron final $v_2$).
\label{fig:fig3_s}}
\end{figure}
Considering that observed pions and protons are made of only light constituent quarks, the difference between their $v_2$ should come from the hadronization process and/or hadronic rescattering. We first study the primordial hadrons right after hadronization, before any decay and hadronic rescatterings take place. Figure~\ref{fig:fig1}(a) shows the $v_2$ of primordial charged pion, kaon, and (anti)proton as a function of $p_{\perp}$ (solid curves). It demonstrates that the mass ordering of $v_{2}$ at low $p_{\perp}$ in AMPT comes from 
the dynamic in coalescence. The dynamical ``selections'' of constituent quarks into pions, kaons, and protons are manifest in the constituent quark $v_2$ distributions shown by the dashed curves in Fig.~\ref{fig:fig1}(a), plotted at the respective {\em hadron} $p_{\perp}$. 

Figure~\ref{fig:fig1}(b) shows the $v_2$ results by $\phi$-randomized AMPT~\cite{He:2015hfa} for primordial hadrons right after coalescence hadronization at the corresponding constituent quark $v_2$'s. No hydrodynamic anisotropic flow is present in the $\phi$-randomized case~\cite{He:2015hfa}, however, mass splitting is still present. It implies that the mass splitting is mainly due to kinematics in the coalescence process. It is therefore not a unique signature of collective anisotropic flow or hydrodynamics.

After hadronization, particles undergo decay and rescattering. We evaluate the $v_2$ of hadrons before hadronic rescattering but including effects of resonance decays by setting the cut-off
time to be 0.6 {\rm fm}/c. The results are shown in Fig.~\ref{fig:fig2}(a) by the dashed curves. The decay product $v_2$ is usually smaller than their parent $v_2$~\cite{Li:2016ubw}. By including decay products, the hadron $v_2$ is reduced from that of primordial hadrons (solid curves in Fig.~\ref{fig:fig1}(a)).

The $v_2$ values before hadronic rescattering (including resonance decay effects) can be considered as the initial $v_2$ for the hadronic evolution stage. The final-stage freezeout hadron $v_2$'s (also including decay daughters) are shown in Fig.~\ref{fig:fig2}(a) by the solid curves. Pion $v_2$ increases while proton $v_2$ decreases from before to after hadronic rescattering. This may be due to interactions between pions and protons, after which they tend to flow together at the same velocity. Thus, the same-velocity pions and protons (i.e.~small $p_{\perp}$ pions and large $p_{\perp}$ protons) will tend to have the same anisotropy. It will yield lower $v_2$ for the protons and higher $v_2$ for the pions at the same $p_{\perp}$ value. This should happen whether or not there is a net gain in the overall charged hadron $v_2$, which depends on the initial configuration geometry from which the hadronic evolution begins~\cite{Li:2016ubw}. The similar results in $d$+Au\ collisions as shown in Fig.~\ref{fig:fig2}(b)~\cite{Li:2016ubw}. 

To summarize the origin of $v_2$ mass splitting, we plot the $v_2$ of pions, kaons, and protons within $0.7<p_{\perp}<0.8$~GeV/$c$, a typical region for different stages of the collision system evolution as shown in Fig.~\ref{fig:fig3_s}. The evolution stages contain:
 (i) right after coalescence hadronization including only primordial hadrons (labeled ``prim.''); 
 (ii) right after coalescence hadronization but including decay products (labeled ``w/ decay''); 
 and (iii) at final freezeout (labeled ``w/ rescatt.~w/ decay). We can see the mass ordering actually comes from interplay of several  physics effects; it comes from coalescence, and more importantly,  from hadronic rescattering process.
\begin{figure}[tbph]
\centering
  \includegraphics[width=0.495\linewidth,height=5cm,clip=]{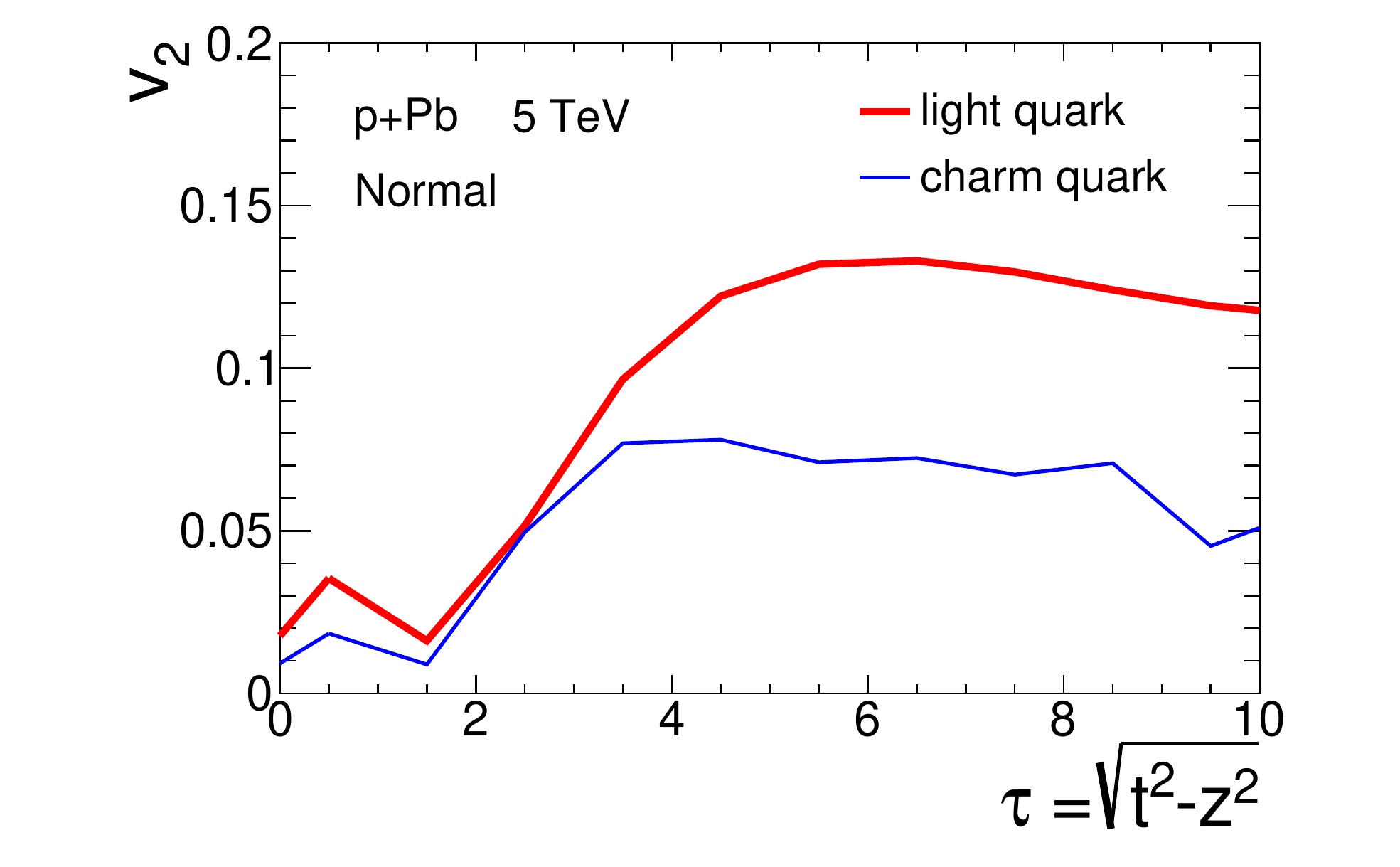}
   \includegraphics[width=0.495\linewidth,height=5cm,clip=]{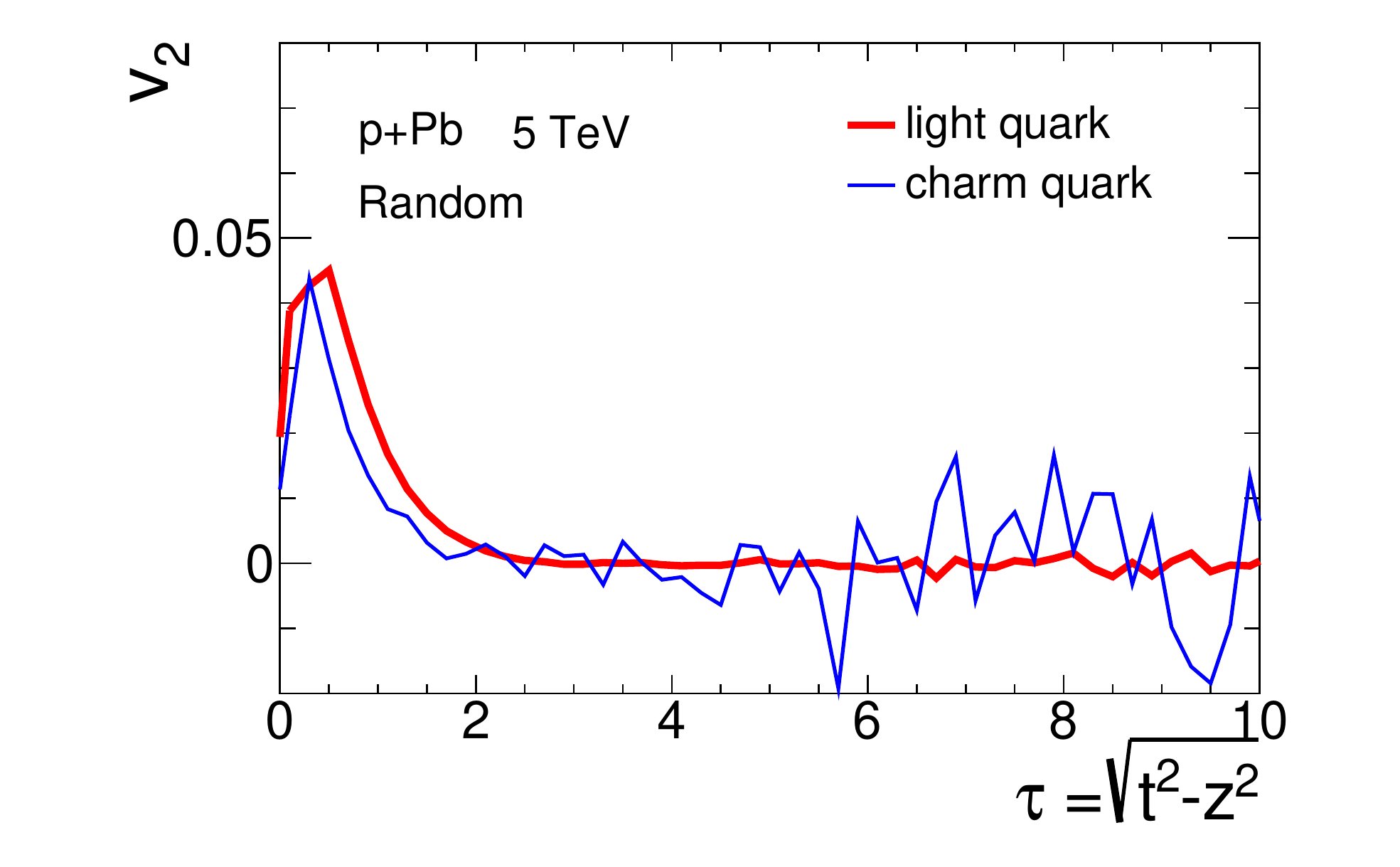}
  \caption{(Color online) The charm and light quark $v_2$ as a function of the proper time $\tau ={\sqrt{t^2-z^2}}$ in p+Pb collisions with {$b=0$~fm}. Both the normal (left) and $\phi$-randomized AMPT (right) results are shown. 
 \label{fig:fig4_c}}
\end{figure}
\section{Charm $v_2$ mechanism\label{sec:setup}}
\quad 
In this section we discuss the production mechanism of charm quark flow in pPb. 
We compare the light and charm quark freeze out $v_2$, integrated over all $p_{\perp}$ at the same proper time. This is shown in the left panel of Fig.~\ref{fig:fig4_c}. The charm $v_{2}$ is systematically lower than light quark $v_{2}$. This may suggest that the charm quark is not as thermalized as light quarks. The right panel of Fig.~\ref{fig:fig4_c} shows $\phi$-randomized results. The charm and light quark $v_2$ are similar suggesting a common escape mechanism.  
\section{Summary\label{sec:summary}}
\quad 
In these proceedings, we investigate the origin of mass splitting of identified hadron $v_{2}$. The mass splitting is due to coalescence and, more importantly,  hadronic rescattering. It is therefore not a unique signature of hydrodynamics. We also study the development of charm $v_{2}$ and found the escape mechanism to be the major contribution in pPb similar to light quarks. 

\ack
This work is supported in part by US~Department of Energy Grant No.~DE-FG02-88ER40412 (LH,FW,WX) and No.~DE-FG02-13ER16413 (DM). HL acknowledges financial support from the China Scholarship Council.

\end{document}